\documentclass[prl,a4paper,10pt,twocolumn,floatfix,superscriptaddress,amsmath,amsfonts,amssymb,longbibliography,citeautoscript]{revtex4-1} 
\usepackage[T1]{fontenc}
\usepackage[utf8]{inputenc}
\usepackage{dcolumn,graphicx,color,booktabs,microtype,afterpage}
\graphicspath{{./}{figure/}}
\usepackage[charter,greekuppercase=italicized]{mathdesign}
\usepackage{lineno}

\makeatletter
\renewcommand\section{\@startsection {section}{1}{\z@}%
   {-2.5ex \@plus -1ex \@minus -.2ex}%
   {0.4ex \@plus.2ex}%
  {\normalfont\large\bfseries}}
\makeatother
\makeatletter
\renewcommand\subsection{%
  \@startsection{subsubsection}{3}{\z@}{3.25ex \@plus1ex \@minus.2ex}%
  {-1em}{\normalfont\normalsize\bfseries}}
\makeatother
\renewcommand{\figurename}{Figure}
\renewcommand{\tablename}{Table}
\makeatletter\renewcommand{\fnum@figure}[1]{\textbf{\figurename~\thefigure~|\ }}\makeatother
\makeatletter\renewcommand{\fnum@table}[1]{\tablename~\thetable.}\makeatother

\newcount\hh \newcount\mm
\hh=\time \divide\hh by 60
\mm=\hh \multiply\mm by 60 \mm=-\mm
\advance\mm by \time
\def\now{\number\hh:\ifnum\mm<10{}0\fi\number\mm}

\newcommand{\citens}[1]{\textsuperscript{\citenum{#1}}}

\newcommand{\LOFP}{La\-Fe\-As$_{1-x}$\-P$_x$\-O}
\begin{document}

\makeatletter\renewcommand{\ps@plain}{%
\def\@evenhead{\hfill\itshape\rightmark}%
\def\@oddhead{\itshape\leftmark\hfill}%
\renewcommand{\@evenfoot}{\hfill\small{--~\thepage~--}\hfill}%
\renewcommand{\@oddfoot}{\hfill\small{--~\thepage~--}\hfill}%
}\makeatother\pagestyle{plain}

\title{Nodal-to-nodeless superconducting order parameter in La\-Fe\-As$_{1-x}$\-P$_x$\-O\\ synthesized under high pressure}

\author{T.\,Shiroka}\email[Corresponding author: \vspace{8pt}]{tshiroka@phys.ethz.ch}
\affiliation{Laboratorium f\"ur Festk\"orperphysik, ETH H\"onggerberg, CH-8093 Z\"urich, Switzerland}
\affiliation{Paul Scherrer Institut, CH-5232 Villigen PSI, Switzerland}

\author{N. Barbero}
\affiliation{Laboratorium f\"ur Festk\"orperphysik, ETH H\"onggerberg, CH-8093 Z\"urich, Switzerland}

\author{R. Khasanov}
\affiliation{Paul Scherrer Institut, CH-5232 Villigen PSI, Switzerland}

\author{N. D. Zhigadlo}
\affiliation{Department of Chemistry and Biochemistry, University of Bern, CH-3012 Bern, Switzerland}

\author{H.-R.\,Ott}
\affiliation{Laboratorium f\"ur Festk\"orperphysik, ETH H\"onggerberg, CH-8093 Z\"urich, Switzerland}
\affiliation{Paul Scherrer Institut, CH-5232 Villigen PSI, Switzerland}

\author{J.\,Mesot}
\affiliation{Laboratorium f\"ur Festk\"orperphysik, ETH H\"onggerberg, CH-8093 Z\"urich, Switzerland}
\affiliation{Paul Scherrer Institut, CH-5232 Villigen PSI, Switzerland}

\begin{abstract}
\noindent 
Similar to chemical doping, pressure produces and stabilizes new phases 
of known materials, whose properties may differ greatly from those 
of their standard counterparts. Here, by considering a series of 
LaFeAs$_{1-x}$P$_x$O iron-pnictides synthesized under high-pressure 
high-temperature conditions, we investigate the simultaneous 
effects of pressure and isoelectronic doping in the 1111 family. 
Results of numerous macro- and microscopic technique measurements, 
unambiguously show a radically different phase diagram for the pressure-grown 
materials, characterized by the lack of magnetic order and the persistence of 
superconductivity across the whole $0.3 \leq x \leq 0.7$ doping range. This 
unexpected scenario is accompanied by a branching in the electronic 
properties across $x = 0.5$, involving both the normal and superconducting phases. 
Most notably, the superconducting order parameter evolves from 
nodal (for $x < 0.5$) to nodeless (for $x \geq 0.5$), in clear 
contrast to other 1111 and 122 iron-based materials grown under ambient-pressure 
conditions.
\end{abstract}

\maketitle\enlargethispage{3pt}

\noindent{}
Superconductivity in LaFePO\citens{Kamihara2006}, a 
compound first synthesized by Zimmer et al.\citens{Zimmer1995}, sets in at 
a modest $T_c$ of only 3.2\,K. However, the significantly higher $T_c = 26$\,K, 
reported later for F-doped LaFeAsO\citens{Kamihara2008}, brought to attention 
a whole new class of compounds, the iron-based layered pnictides and 
chalcogenides, whose complex magnetic and superconducting properties are 
still being investigated\citens{Johnston2010,Paglione2010,Si2016}.
Although the electronic spin fluctuations are widely acknowledged as 
responsible for the pairing mechanism in the superconducting phase\citens{Inosov2016}, 
many issues still remain open\citens{Borisenko2016}. 
For instance, surprisingly, two rather similar, isostructural 
and isovalent 1111 compounds, such as LaFePO and LaFeAsO, 
exhibit strikingly different properties. 
While the first is paramagnetic and becomes superconducting 
below 5\,K\citens{Kamihara2006,Hamlin2008} (with indications 
that oxygen vacancies might also influence 
$T_c$)\citens{McQueen2008,Analytis2008}, the second 
compound orders antiferromagnetically below 
$T_\mathrm{N} = 140$\,K\citens{Cruz2008}, with no traces 
of superconductivity at lower temperatures.

Due to initial difficulties in preparing high-quality 1111 materials, 
this puzzling behavior attracted first only the attention of theorists.
By means of ab-initio den\-si\-ty-func\-tion\-al methods, the electronic 
structures and the magnetic properties of LaFePO and LaFeAsO were 
calculated in considerable detail\citens{Singh2008,Lebegue2007}. 
It turned out that pnictogen atoms play a key role in establishing 
the Fe-P (or Fe-As) distance, giving rise to an unusual sensitivity 
of material's properties to an apparently minor 
detail\citens{Lebegue2009}.
This conclusion was reinforced by later work, where an interpretation 
based on quantum criticality (QC) was put forward\citens{Abrahams2011}. 
In a QC scenario, the proximity of iron-based materials to a Mott 
transition implies that, by increasing the ratio of kinetic energy to 
Coulomb repulsion, one can pass from an antiferro- to a paramagnetic 
state. 
Detailed calculations in the related F-doped LaFeAsO materials 
showed the proximity of the latter to a quantum tricritical point, 
with an anomalously flat energy landscape, implying that even weak 
perturbations can induce significant changes in the physical 
properties\citens{Giovannetti2011}. 
Magnetic frustration is believed to cause such behavior, since the 
large degeneracy of the ground state close to a quantum critical 
point (QCP), (i.e., entropy accumulation) can be relieved by a 
low-temperature transition to the superconducting state\citens{Si2016}.

In LaFeAsO, the most obvious way to induce such a quantum-critical 
transition is the isoelectronic substitution of phosphorus for arsenic. 
Indeed, the smaller 
ionic radius of phosphorus leads to a smaller cell volume and, hence, to 
an enhanced kinetic energy and to reduced electronic correlations. Amid 
the antiferro- and paramagnetic behavior of the pristine As and 
P compounds, respectively, one expects a superconducting dome, with 
the highest $T_c$ being reached at the QCP\citens{Abrahams2011}.

These predictions were first tested in a systematic study of the 
La\-Fe\-As$_{1-x}$\-P$_x$\-O series, which focused on x-ray 
structural analysis, bulk resistivity, and magnetometry 
measurements\citens{Wang2009}. 
By partially substituting P for As, the Fe$_2$As$_2$ 
layers were reported to contract, while the La$_2$O$_2$ 
layers to expand along the $c$-axis. Superconductivity (SC) 
occurred in a narrow range around $x = 0.3$, with a rather low 
maximum $T_c$ of 10\,K. The absence of superconductivity 
above $x = 0.4$, yet its reappearance in LaFePO, i.e., for $x = 1$, 
remained an open issue. No experimental evidence indicating 
the occurrence of a QCP at $x = 0.3$ was found.
On the other hand, the As-for-P substitution in 122 systems, such as 
BaFe$_2$(As$_{1-x}$P$_x$)$_2$, showed that the AFM phase at 
$x = 0$ was gradually replaced by a superconducting phase at $x = 1$, 
with a putative QCP occurring at $x = 0.3$.\citens{Kasahara2010}

More recent efforts included microscopic investigations of 
the La\-Fe\-As$_{1-x}$\-P$_x$\-O series via ${}^{31}$P 
nuclear magnetic resonance (NMR)\citens{Kitagawa2014}.
In this case, resonance-width data suggested the onset of 
antiferromagnetism in different ranges of $x$ substitutions, 
with the resulting phase diagram not showing a clearcut QCP, 
but rather AF zones separated by SC ``pockets''. 
Very recently, similar  SC ``pockets'' were also found in the rather 
complex hole- and electron-doped (La,Sr)FeAs$_{1-x}$P$_{x}$(O,F/H) 
system\citens{Miyasaka2017}.

To address the many issues mentioned above, such as the reasons 
for the very different electronic properties of LaFeAsO and LaFePO, 
the unusual sensitivity to structural modifications, and the occurrence 
of quantum criticality, we investigated a new batch of 
La\-Fe\-As$_{1-x}$\-P$_x$\-O compounds, grown under high-pressure 
conditions. These conditions are known to stabilize otherwise unstable 
(or energetically unfavorable) phases and allowed us to study the 
consequences of the \textit{simultaneous} occurrence of chemical- 
(via substitution) and physical (during synthesis) pressure. 
As we show here, the latter leads to surprising results in the 1111 
class. Thus, by employing local microscopic techniques, such as 
muon-spin rotation ($\mu$SR) and nuclear magnetic resonance (NMR), 
we obtain a radically revised low-temperature La\-Fe\-As$_{1-x}$\-P$_x$\-O 
phase diagram, characterized by the lack of antiferromagnetic transitions 
at intermediate $x$ values (between 0.3 and 0.7). 
In addition, on the basis of new data, we bring new evidence about 
the interplay of magnetic fluctuations and superconductivity.
Most importantly, coherent experimental results indicate a clear 
change in the character of the superconducting order parameter, 
which appears to evolve from nodal to nodeless as $x$ increases, 
the exact opposite with respect to standard ambient-pressure grown 
samples\citens{Hashimoto2012,Nourafkan2016}.

\section{\label{sec:results}Results}
\subsection{\label{sec:xray_magnetism}Structural, magnetic, and 
transport properties.} 
The x-ray powder diffraction patterns of \LOFP\ are shown in 
the Supplementary Fig.~\ref{fig:diffraction} and confirm that 
the studied compounds adopt the expected overall structure. Indeed, 
our specimens, grown via high-pressure synthesis, reveal 
diffraction patterns that are almost indistinguishable from those 
of samples grown under standard conditions\citens{Wang2009}. Yet, 
in detail the evolution of the multiple peaks close to 30 
degrees with $x$ is different in our case, indicating different 
local environments. As we show below, this leads to a radically different 
phase diagram and SC properties. The tetragonal ($P4/nmm$) crystal 
structure of \LOFP\ evolves smoothly from $a = 4.03$\,\AA\ and 
$c =  8.72$\,\AA\ for $x = 0$ to $a =  3.96$\,\AA\ and $c =  8.51$\,\AA\ 
for $x = 1$, the decrease in lattice parameters reflecting the smaller 
ionic radius of P with respect to As. The absence of substantial 
structural differences between samples of this series indicates that the 
observed changes in the electronic properties and, hence, the 
adopted ordered phases at low temperatures, are related to 
electron-correlation effects. How these tiny structural differences 
cause the alleged variation in electron correlations 
is the challenging task for future refined studies.

The superconducting critical temperatures $T_c$ were determined by means of 
SQUID magnetometry and radio-frequency detuning of the NMR resonant 
circuit (see Supplementary Fig.~\ref{fig:magnetization}), 
with all samples exhibiting large fractions of magnetic shielding and the 
maximum $T_c$ being reached at $x = 0.5$ (see Fig.~\ref{fig:phase_diag}).
This is a surprising result, clearly departing from known phase diagrams 
of La-1111 samples grown at ambient pressure\citens{Mukuda2014}, 
for which no superconductivity is observed in the $x=0.4$ to 0.7 range.
In our case, low-temperature, low-field susceptibility data show a 
relatively steep decrease of $\chi(T)$ below $T_c $ and a significant 
diamagnetic response close to $T = 0$, indicating a good chemical 
homogeneity and bulk superconductivity, respectively. 
From the depression of  $T_c$ with increasing magnetic fields we 
estimate an $H_{c2}(0) \sim 70$\,T, a value that matches data 
reported in the literature for various  La-1111 compounds\citens{Prakash2008,Singh2012}.

\begin{figure}[t]
\centering
\includegraphics[width=0.85\columnwidth]{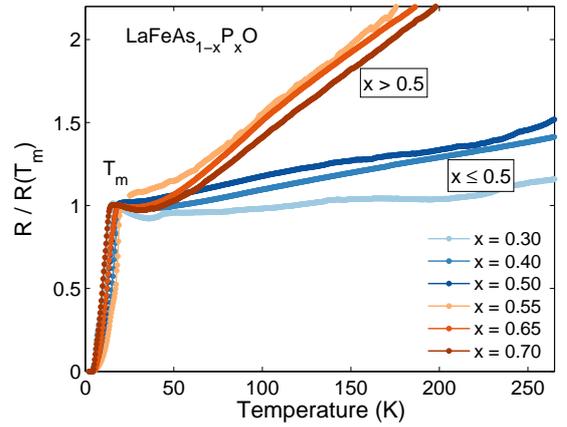}
\caption{\label{fig:resistivity}\textbf{Split resistivity curves.} 
Normalized resistivity vs.\ temperature 
for the LaFeAs$_{1-x}$P$_x$O series synthesized under high-pressure 
conditions. All samples become superconducting after reaching a local 
maximum at $T_m$. Notice the different behavior of samples with $x \leq 0.5$ 
from those with $x > 0.5$, highlighted by two different color hues.}
\end{figure}

The temperature dependence of resistivity $\rho(T)$ is shown in 
Fig.~\ref{fig:resistivity}. Unlike previously reported results 
(see, e.g., Ref.~\onlinecite{Wang2009}), \textit{all} our (high-pressure 
grown) samples are superconductors with $T_{c}$ values in the 
15--20\,K range. Likewise, all of them exhibit a shallow maximum 
 at $T_{m}$, just above the superconducting transition, related to increased 
electronic correlations (see below). 
By normalizing $\rho(T)$ to the peak occurring at $T_{m}$ 
[and not to the usual $\rho$(300\,K) value], we find an intriguing 
splitting into two branches. Samples with $x \leq 0.5$ show a rather weak 
temperature dependence and aggregate into the lower branch, while 
those with $x > 0.5$ exhibit a stronger $T$-dependence and 
populate the upper branch. This is a remarkable result, indicating a 
profound change in the electronic correlations across the $x = 0.5$ 
boundary, confirmed also by microscopic probes (see next sections). 
Note that, by plotting existing data\citens{Wang2009} 
in the same way produces only uniformly spaced curves, thus 
indicating the particular nature of the high-pressure grown samples.

\subsection{\label{sec:muSR_results}Absence of magnetic order 
from zero-field $\mu$SR.} To reveal the magnetic and superconducting behavior 
of the \LOFP\ series, we investigated systematically the temperature dependence 
of the muon-spin relaxation in zero- and in applied magnetic fields, respectively.
As a local microscopic technique, muon-spin rotation/relaxation ($\mu$SR), relies on 
the detection of muon-decay positrons, emitted preferentially along the muon-spin 
direction\citens{Blundell1999,Yaouanc2011}. 
Given the absence of perturbing applied fields, zero-field (ZF) $\mu$SR 
represents a uniquely sensitive probe of the intrinsic magnetic properties, in many 
respects complementary to NMR/NQR.

\begin{figure*}[t]
\centering
\hspace{2mm}\includegraphics[width=.72\columnwidth]{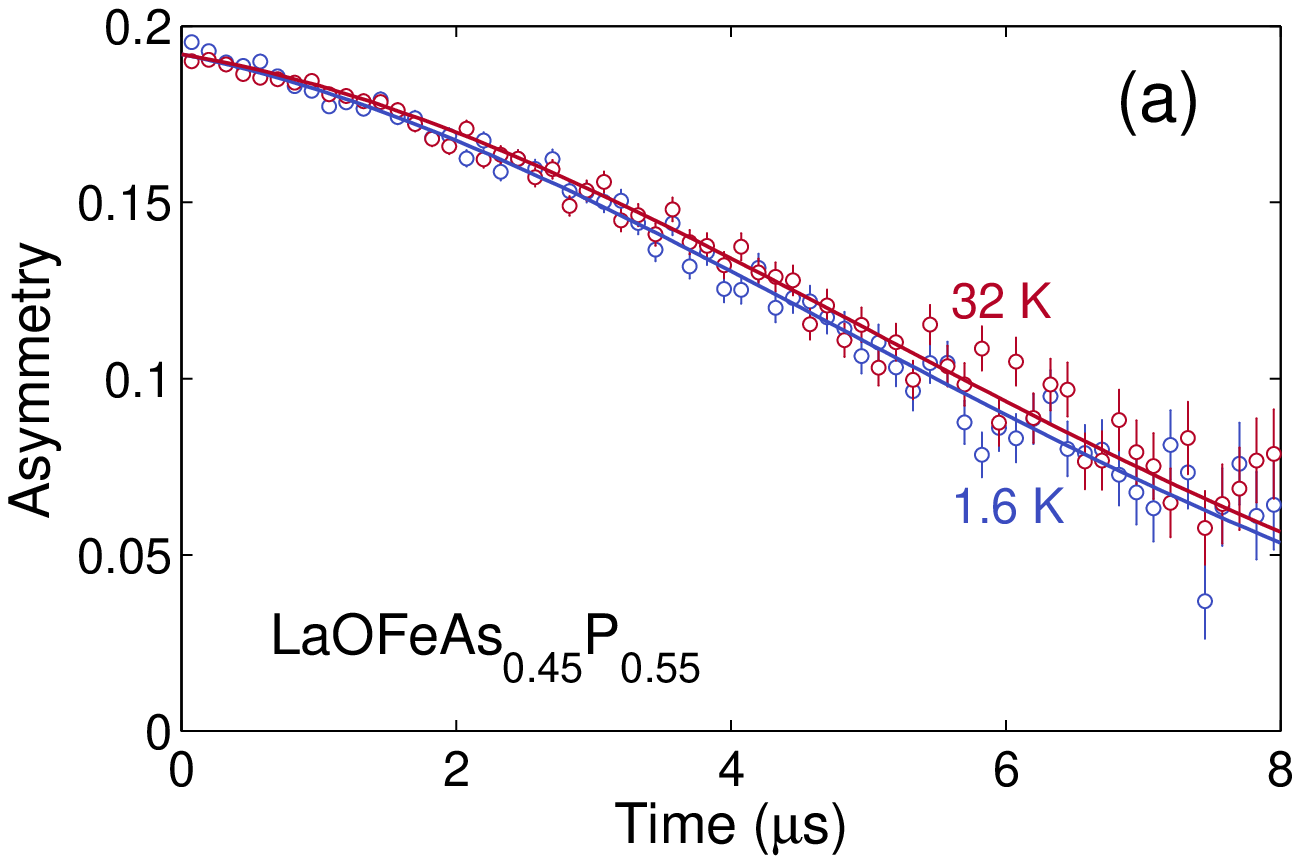} 
\hspace{6mm}\includegraphics[width=0.5\columnwidth]{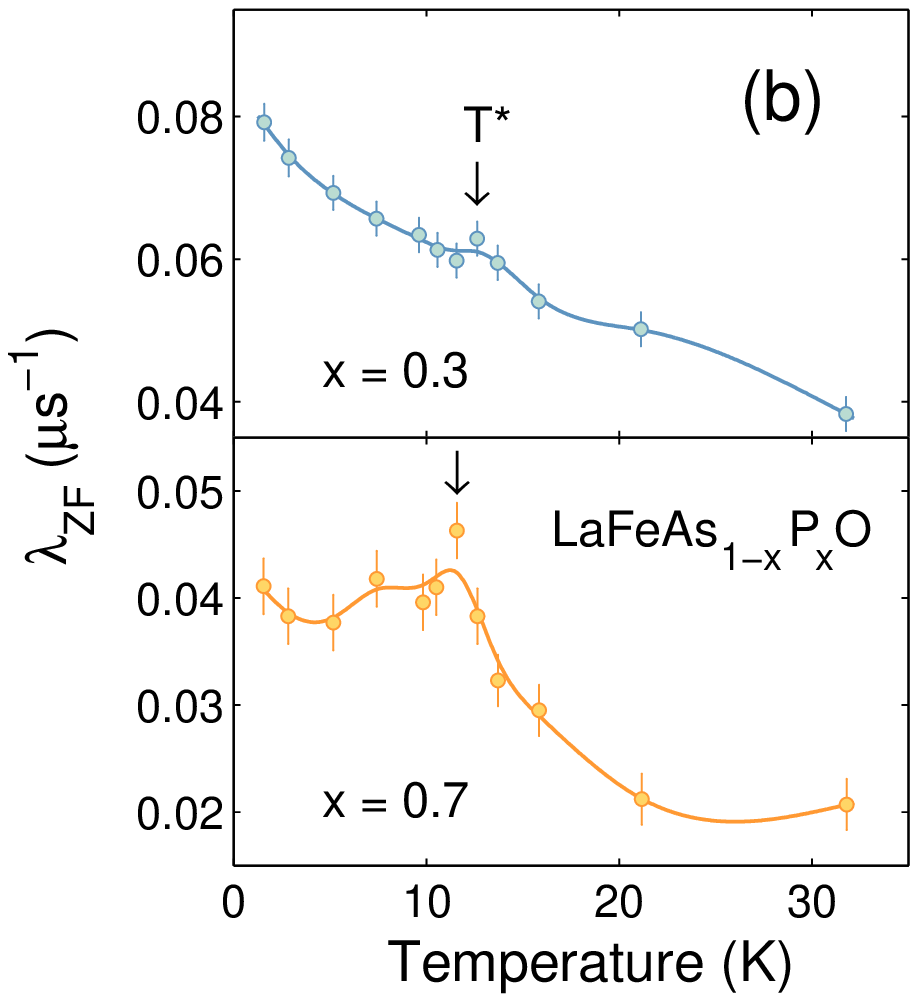}
\caption{\textbf{Zero-field $\boldsymbol{\mu}$SR relaxation.}
(a) Representative LaFeAs$_{1-x}$P$_x$O zero-field $\mu$SR spectra 
above and below $T_c$, for $x = 0.55$, fitted by means of Eq.~(\ref{eq:ZF_muSR}). 
(b) For many samples the relaxation is rather inconspicuous, 
yet it invariably shows a tiny peak close to $T^*$ (arrows), corresponding to a 
maximum in electronic spin fluctuations (see text). Lines are guides to the eye. 
}
\label{fig:ZF-MuSR}
\end{figure*}

\begin{figure*}[ht]
\includegraphics[width=0.75\columnwidth]{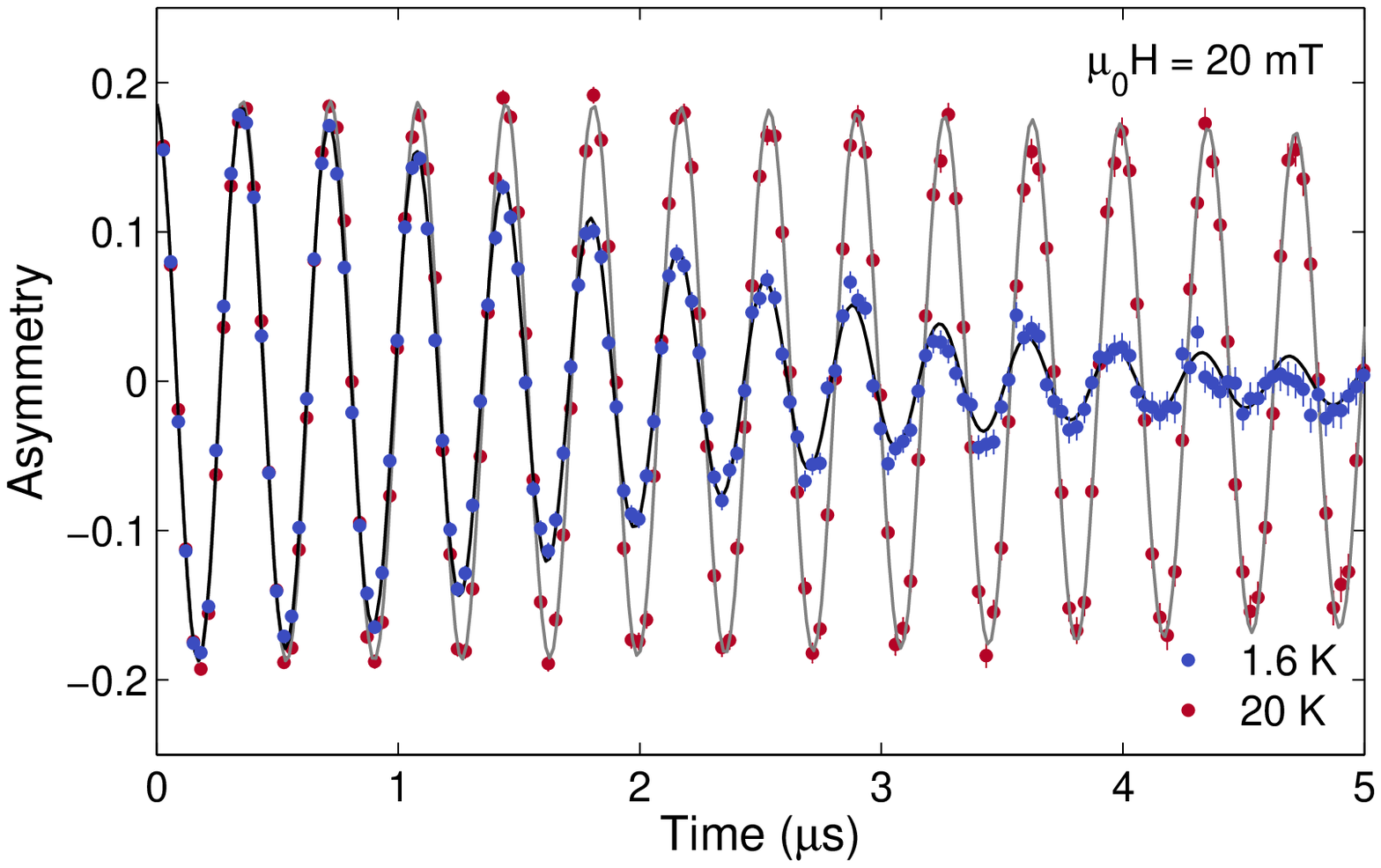}
\hspace{3mm}\includegraphics[width=0.584\columnwidth]{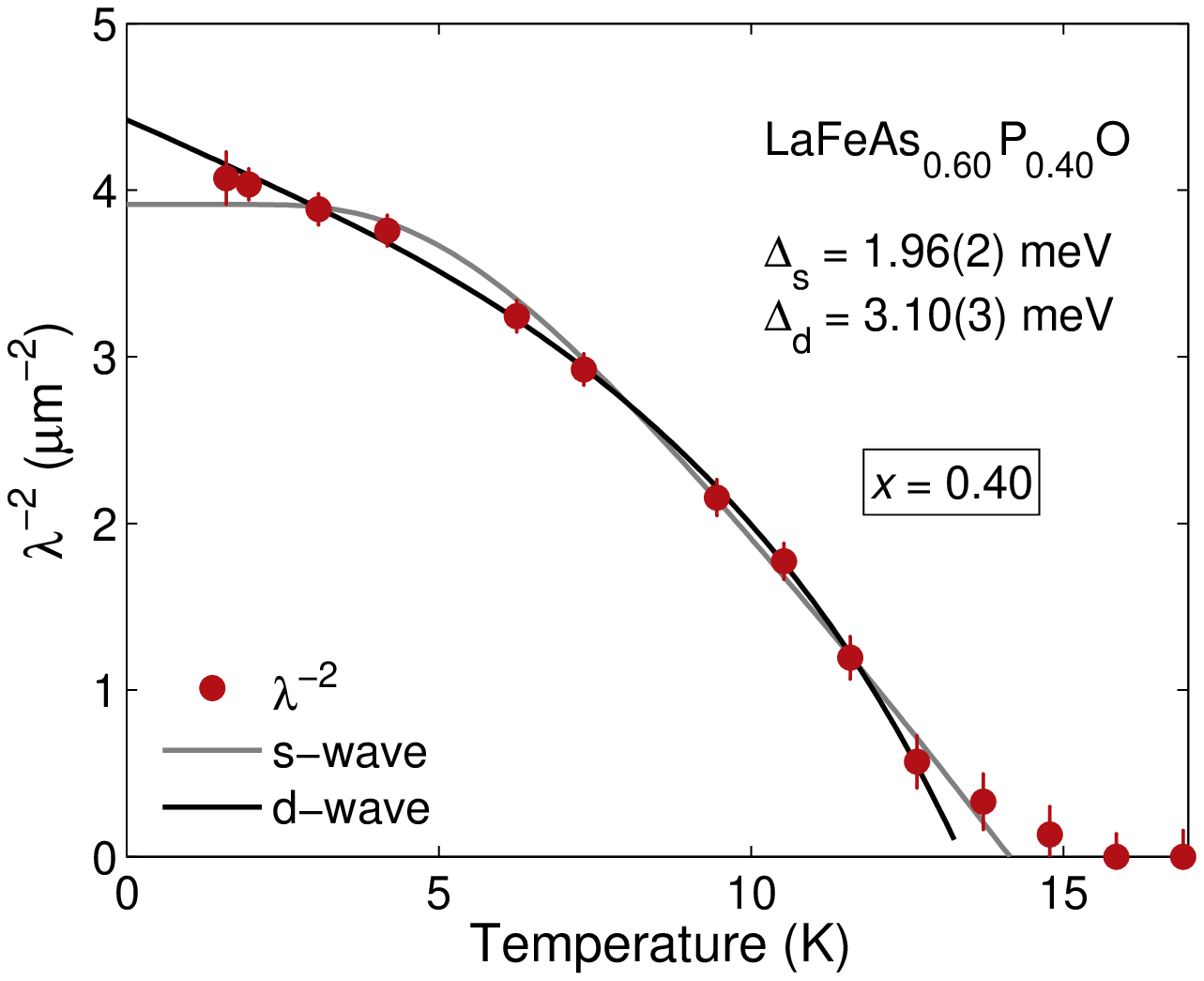}
\hspace{1mm}\includegraphics[width=0.580\columnwidth]{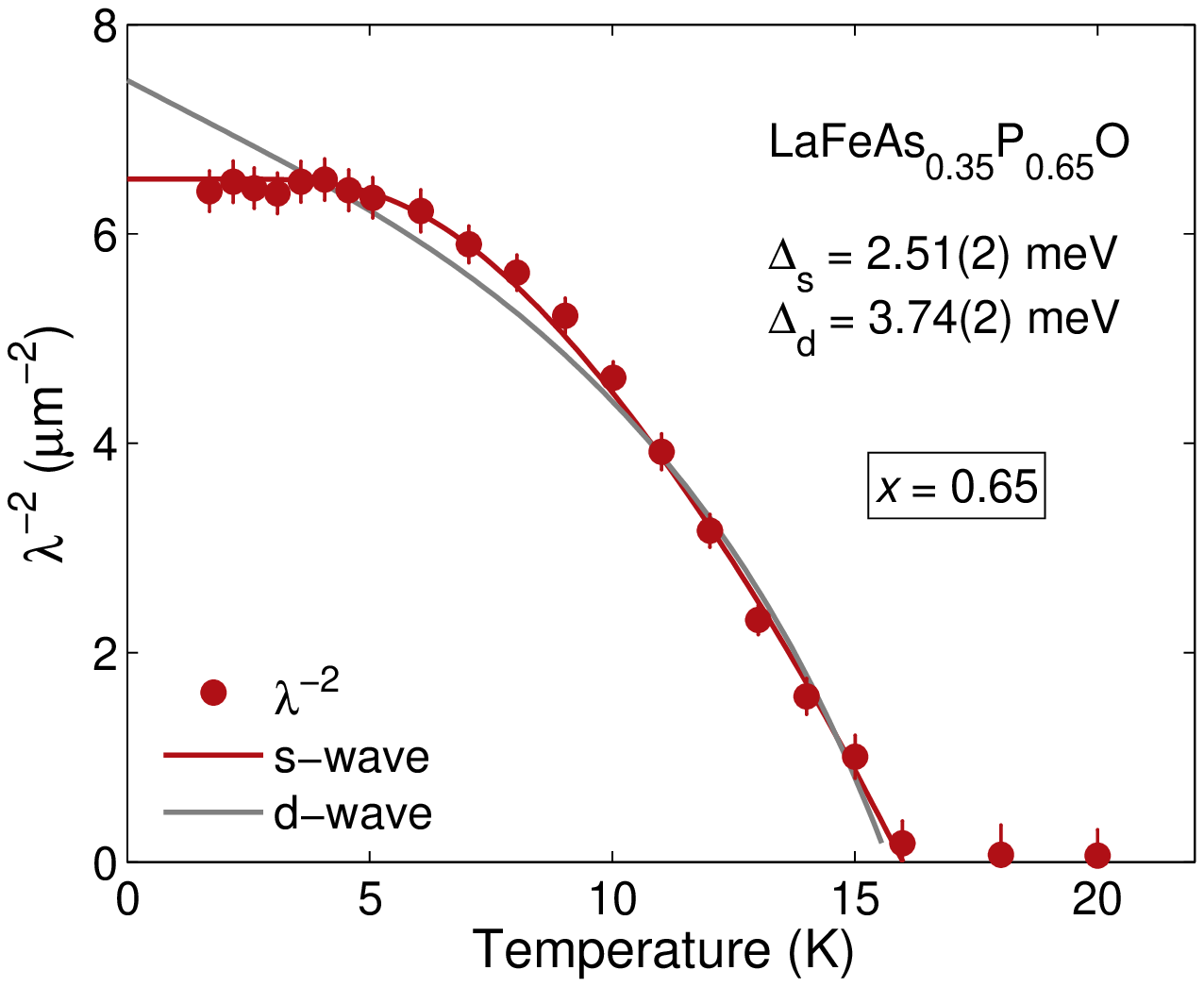}
\caption{\label{fig:TF-MuSR}\textbf{Changing nature of SC from 
TF-$\boldsymbol{\mu}$SR data.}
(a) Representative LaFeAs$_{1-x}$P$_x$O $\mu$SR 
spectra for $x = 0.7$, taken above and below $T_{c}$ in a 20-mT 
transverse field. The onset of superconductivity upon lowering the 
temperature is reflected in a faster decay of the asymmetry.
Temperature dependence of 
$\lambda^{-2}$ for $x = 0.4$ (b) and $x = 0.65$ (c) measured at 
$\mu_{0}H = 20$\,mT. Solid lines indicate fits by means of $s$- and $d$-wave SC pairings.
Notice the change in pairing character across $x = 0.5$. See text for details.
}
\end{figure*}

Typical ZF-$\mu$SR data for the $x = 0.55$ case are shown in Fig.~\ref{fig:ZF-MuSR}a. 
The $\mu$SR asymmetry spectra at 32 and 1.6\,K, i.e., above and below $T_c$ and/or 
a possible magnetic ordering temperature $T^{*}$, do not exhibit any oscillations, 
but only a weak decay, best described by a Kubo-Toyabe relaxation function\citens{Yaouanc2011} 
multiplied by an exponential decay:
\begin{equation}
\label{eq:ZF_muSR}
A_\mathrm{ZF} = A_0 %
\left[\frac{1}{3} + \frac{2}{3}(1- a^2t^2)\exp{\left(\frac{-a^2t^2}{2}\right)} \right] \exp(-\lambda_\mathrm{ZF} t).
\end{equation}
Here $A_0$ is the initial asymmetry parameter, while $a$ and $\lambda_\mathrm{ZF}$ are 
the muon-spin relaxation rates due to static nuclear moments and electronic 
moments, respectively.
The nuclear contribution is small, almost temperature independent, and accounts 
for the initial Gaussian-like decay.
Hence, the observed depolarization is mostly determined by contributions 
from the electronic magnetic moments. The key feature of the data shown 
in Fig.~\ref{fig:ZF-MuSR}a is the \emph{unchanged relaxation} rate above and 
below $T_c$ ($T^*$). 
This is remarkable since, at low temperatures, most 
iron-based superconductors exhibit antiferromagnetic order which, 
depending on whether long- or short-ranged, implies either muon-spin 
asymmetry oscillations or a strong increase in damping, respectively 
(see, e.g., Refs.~\onlinecite{Luetkens2009,Drew2009}). 
The absence of either of them in the investigated \LOFP\ series 
rules out the onset of a possible magnetic order, in clear contrast 
with other cases, where a magnetic order (long- or short-ranged) is 
established, alone or in coexistence with 
superconductivity\citens{Shiroka2011,Lamura2015}.
Yet, as shown in  Fig.~\ref{fig:ZF-MuSR}b, the relaxation rates still 
exhibit a small hump close to $T^*$ (corresponding to a maximum in the 
electronic spin fluctuations), showing up prominently in the NMR relaxation data 
(see Fig.~\ref{fig:invT1T_shift}a).
The origin of the hump relates to the competing SC and magnetic order 
in superconductors with $s^{\pm}$ pairing, which below $T_{c}$ tends to 
suppress the magnetically induced increase in relaxation rate, thus 
giving rise to a cusp in the relaxation data.\citens{Fernandes2010}

\subsection{Change of SC pairing characteristics revealed via TF-$\mu$SR.}
Transverse-field (TF) $\mu$SR is among the standard techniques for 
studying the superconducting phase. When an external magnetic field is 
applied to a field-cooled type-II superconductor, the resulting flux-line 
lattice (FLL) modulates the local field. Implanted muons sense uniformly 
the SC-related field inhomogeneity, which is 
detected as an additional Gaussian relaxation $\sigma_{\mathrm{sc}}$. 
Figure~\ref{fig:TF-MuSR} clearly illustrates this by means of typical 
TF-$\mu$SR spectra for $x = 0.7$, measured at $\mu_0H = 20$\,mT, both 
above and below $T_c$ (15\,K). As the temperature is lowered below $T_c$, 
the asymmetry relaxation rate increases significantly. In the TF-$\mu$SR 
case, the time-domain $\mu$SR data were fitted using\citens{Yaouanc2011}:
\begin{equation}
A_\mathrm{TF} = A_\mathrm{TF}(0) \, \cos(\gamma_{\mu} B_{\mu} t + \phi) 
e^{-0.56\lambda_\mathrm{ZF} t} e^{- \sigma^2 t^2/2}.
\end{equation}
Here $A_\mathrm{TF}(0)$ is the initial asymmetry, 
$\gamma_{\mu} \-= \-2\pi \times 135.53$\,MHz/T
is the muon gyromagnetic ratio, $B_{\mu}$ is the local field at the 
implanted-muon site, $\phi$ is the initial phase, and $\lambda_\mathrm{ZF}$ 
and  $\sigma$ are an exponential and a Gaussian relaxation rate, respectively. 
The weak exponential relaxation $0.56\lambda_\mathrm{ZF}$\citens{Yaouanc2011} 
was chosen in agreement with the ZF data analysis and is considerably smaller 
than the Gaussian relaxation rate $\sigma$. The latter 
contains contributions from both the FLL ($\sigma_\mathrm{sc}$) and a small temperature-independent 
relaxation due to nuclear moments ($\sigma_\mathrm{n}$). The FLL contribution below $T_c$ 
was derived by subtracting the nuclear contribution from the Gaussian relaxation rate, i.e., 
$\sigma_\mathrm{sc}^2 = \sigma^2 - \sigma_\mathrm{n}^2$, where $\sigma_\mathrm{n}$ 
was kept fixed at its value above $T_c$.
In all cases we observe a clear diamagnetic shift in the superconducting 
phase, determined as the difference between the applied and the sensed 
magnetic fields. This can also be seen directly in Figure~\ref{fig:TF-MuSR}, 
where at long times the low-temperature oscillations show a reduced frequency. 
Besides diamagnetism, the development of a flux-line lattice below $T_c$ 
implies the appearance of $\sigma_\mathrm{sc}$, in turn reflecting
the increase in $1/\lambda^2$ [see Fig.~\ref{fig:TF-MuSR}], the two 
being related by\citens{Barford1988,Brandt2003}:
\begin{equation}
\label{eq:lambda}
\frac{\sigma_{\mathrm{sc}}^2}{\gamma^2_{\mu}}=0.00371 \cdot \frac{\varPhi_0^2}{\lambda^4},
\end{equation}
with $\varPhi_0 =2.068\times10^{-3}$\,T\,$\mu$m$^2$ the magnetic-flux quantum 
and $\lambda \equiv \lambda_{\mathrm{eff}}$ the effective magnetic-field penetration depth.
In anisotropic polycrystalline superconducting samples (as is the case 
for \LOFP) the effective penetration depth is determined mostly by the 
shortest penetration depth $\lambda_{ab}$, the relation between the two 
being $\lambda_{\mathrm{eff}} = 3^{1/4}\lambda_{ab}$\citens{Fesenko1991}.

Figure~\ref{fig:TF-MuSR} (b and c) shows the temperature dependence of 
$\lambda^{-2}(T)$, proportional to the effective superfluid density $ \lambda^{-2} \propto \rho_s $,
for two representative samples, $x = 0.4$ and $0.65$. In the 
latter case $\lambda^{-2}(T)$ is clearly constant at low temperatures, (below $T_c/3$), 
hence indicating a fully-gapped superconductor (i.e., one with a nodeless SC gap). 
Conversely, the $x = 0.4$ sample, which does not exhibit any saturation of 
$\lambda^{-2}$, even close to $T=0$\,K, behaves as a 
superconductor with an anisotropic (nodal) gap (most probably of $d$ type).
This remarkable \textit{change in the symmetry of pairing} in the superconducting phase, seems to reflect the 
diverse normal-state properties of samples across the $x = 0.5$ composition, 
as already determined from 
resistivity measurements (see Fig.~\ref{fig:resistivity}).
Indeed, the experimental $\lambda^{-2}(T)$ values could only be fitted 
by mutually exclusive $s$- or $d$-wave models which, as shown in 
Fig.~\ref{fig:TF-MuSR}, provide $\Delta_{d}(0) = 3.10(3)$\,meV 
and $\Delta_{s}(0) = 2.51(2)$\,meV for the $x = 0.4$ and 0.65 case, 
respectively (fit details are reported in the appendix).

\vspace{-5pt}
\subsection{\label{sec:NMR_results}NMR line shapes confirm lack of magnetic order.}
Nuclear magnetic resonance (NMR) is a powerful yet complementary technique 
to $\mu$SR, with respect to probe location, presence of polarizing fields, 
time scale, etc. 
By using mostly ${}^{31}$P-NMR measurements, we investigate both the 
static (line widths and -shifts) as well as the dynamic (spin-lattice relaxation) 
properties of the \LOFP\ series. 

\begin{figure}[thb]
\includegraphics[width=0.8\columnwidth]{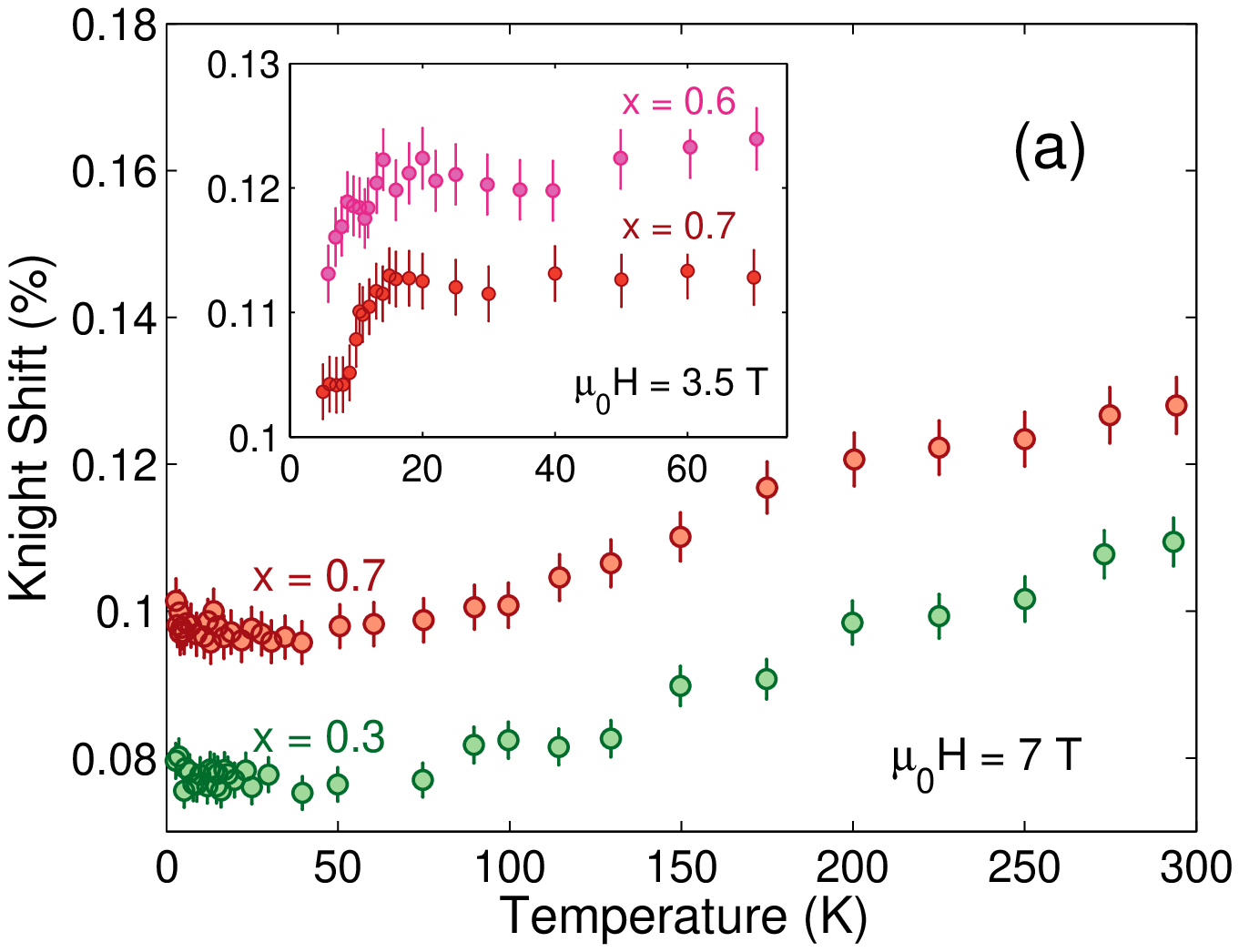}
\vspace*{5mm}
\includegraphics[width=0.8\columnwidth]{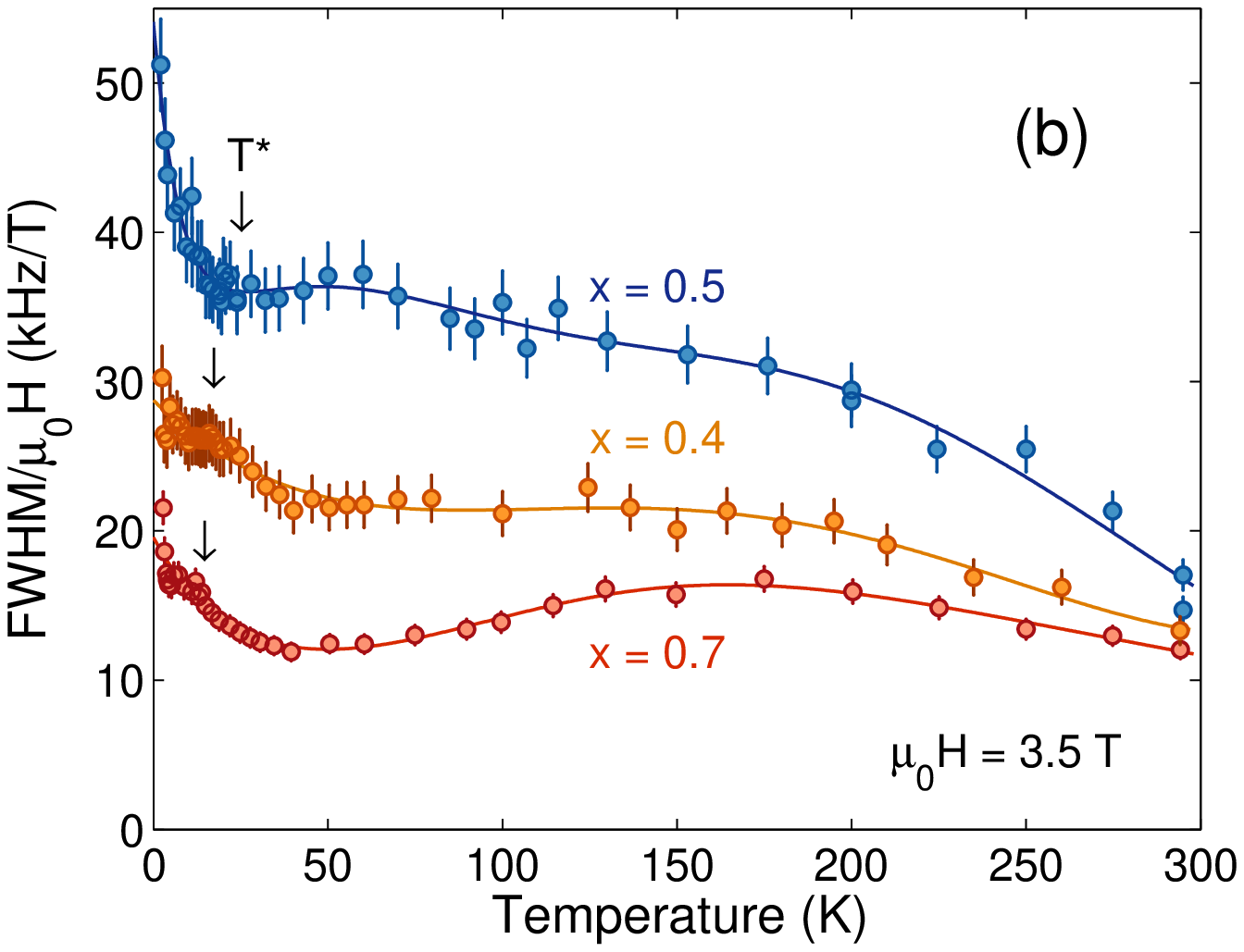}
\caption{\textbf{Weakly increasing linewidths reflect lack of magnetic order.} 
Representative ${}^{31}$P NMR shifts (a) and line widths (b) 
vs.\ temperature, measured at 3.5 and 7\,T. 
The full lines are guides to the eye. A drop in shift below $T_c$ 
(arrow) is observed only at 3.5\,T (inset). The line widths show 
only a moderate increase below $T^{*}$ for all $x$ values 
and applied fields.}
\label{fig:shifts}
\end{figure}

In all cases the ${}^{31}$P NMR lines are narrow (about 20\,kHz) and 
evolve smoothly with temperature (a typical dataset is shown in 
Supplementary Fig.~\ref{fig:lines}).
Given the powder nature of the samples, a linewidth of only 160\,ppm 
indicates a good crystalline quality.
An analysis of line shifts and widths for various samples and applied 
fields reveals a number of interesting features (see Fig.~\ref{fig:shifts}).

The Knight shift, which probes the intrinsic uniform susceptibility, is defined 
as $K = (f_\mathrm{r}-f_0)/f_{0}$, with $f_{0}$ the reference frequency 
of the bare nucleus in an applied field $\mu_{0}H$ and $f_\mathrm{r}$ 
the observed NMR frequency. In our case, the average $K$ values are 
$\sim 0.1$\%, with $K(T)$ decreasing upon reducing the temperature 
and a trend to saturation below 50\,K. 
Significantly enhanced $K(T)$ values, with a maximum at ca.\ 125\,K 
were previously reported in similar compounds, but synthesized at ambient 
pressure (e.g., for $x = 0.7$)\citens{Mukuda2014}. 
Such only partial agreement with our ${}^{31}$P NMR results most likely 
reflects the different sample-synthesis conditions.
The datasets collected at 3.5\,T (see inset), show 
an additional drop in Knight shift upon entering the superconducting phase in 
two representative cases, $x = 0.6$ and 0.7.
Besides being compatible with the $s$-wave nature of superconductivity 
in the \LOFP\ family, this last feature, missing in both our high-field dataset 
as well as in those reported in the literature\citens{Mukuda2014} (taken at 12\,T), 
suggests an active role of the applied field, consisting not merely in 
the well-known lowering of $T_{c}$.
A final interesting feature of the reported Knight-shift data is a 
temperature-independent offset between the $x = 0.3$ and the 0.7 
datasets (main panel in Fig.~\ref{fig:shifts}a). 
The overall decrease in $K(T)$ for $x = 0.3$ corresponds to a reduction 
of the uniform spin susceptibility and is compatible with enhanced 
antiferromagnetic correlations, tending towards the AF order, as 
observed in the $x = 0$ case.
Incidentally, given the symmetric compositions (with respect to $x = 0.5$) 
of the $x=0.3$ and 0.7 compounds, their non-overlapping $K(T)$ curves 
suggest a different strength/nature of electronic correlations, 
above and below $x = 0.5$, as we discuss below.

The linewidth data, reported in Fig.~\ref{fig:shifts}b, are also quite 
informative. In general, samples with $x = 0.5$ or close to it  
exhibit the largest linewidths, compatible with an enhanced degree of 
disorder\citens{Shiroka2011a}.
The increase in FWHM with decreasing temperature --- 
often an indication of a possible magnetic order --- in our case is smooth, 
with only a minor enhancement at the lowest temperatures (as identified 
by arrows in Fig.~\ref{fig:shifts}b). 
This behavior is in good agreement with our ZF-$\mu$SR data, 
showing only minor changes in the relaxation rate across a presumed 
$T_\mathrm{N}$ (see Fig.~\ref{fig:ZF-MuSR}b). At the same time, 
our FWHM data are in stark contrast 
with those of samples synthesized at ambient pressure\citens{Mukuda2014}, 
where a tenfold (or higher) increase in linewidth is observed upon 
entering the antiferromagnetic phase. 
The lack of appreciable variations of FWHM vs.\ $T$ strongly suggests 
that samples synthesized under high-pressure \emph{do not 
exhibit any AF order} at intermediate $x$ values but, as we show below, at 
most sustain (significant) AF fluctuations.

\subsection{NMR relaxation rates and AF spin fluctuations.}
\begin{figure}[thb]
\includegraphics[width=0.9\columnwidth]{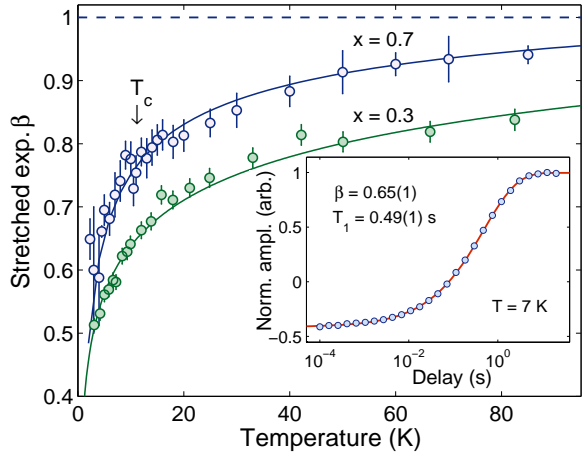}
\caption{\textbf{The simultaneous presence of As and P atoms implies a strongly 
stretched NMR relaxation.} The stretched-exponential coefficient $\beta$ shows a 
monotonic decrease as the temperature is lowered, starting well above $T_c$, 
yet distinct for samples with $x$ values above and below $x = 0.5$.
Inset: the recovery of magnetization in a typical ${}^{31}$P NMR spin-lattice 
relaxation experiment below $T_c$ spans several decades.}
\label{fig:T1_curves_2T}
\end{figure}

The ${}^{31}$P spin-lattice relaxation times $T_1$ were evaluated from 
magnetization-recovery curves $M_z(t)$, such as those shown in the inset 
of Fig.~\ref{fig:T1_curves_2T}, by using the standard expression for the 
exponential recovery of spin-$1/2$ nuclei. For the central transition 
of the spin-$3/2$ ${}^{75}$As nuclei we use\citens{Mcdowell1995}:
\begin{equation*}
\label{eq:T1-IR}
M_z(t) = M_z^0 \left[1 - f (0.9\exp(-6t/T_1)^{\beta} + 0.1\exp(- t/T_1)^{\beta})\right].
\end{equation*}
\begin{figure*}[bht]
\centering
\includegraphics[width=0.85\columnwidth]{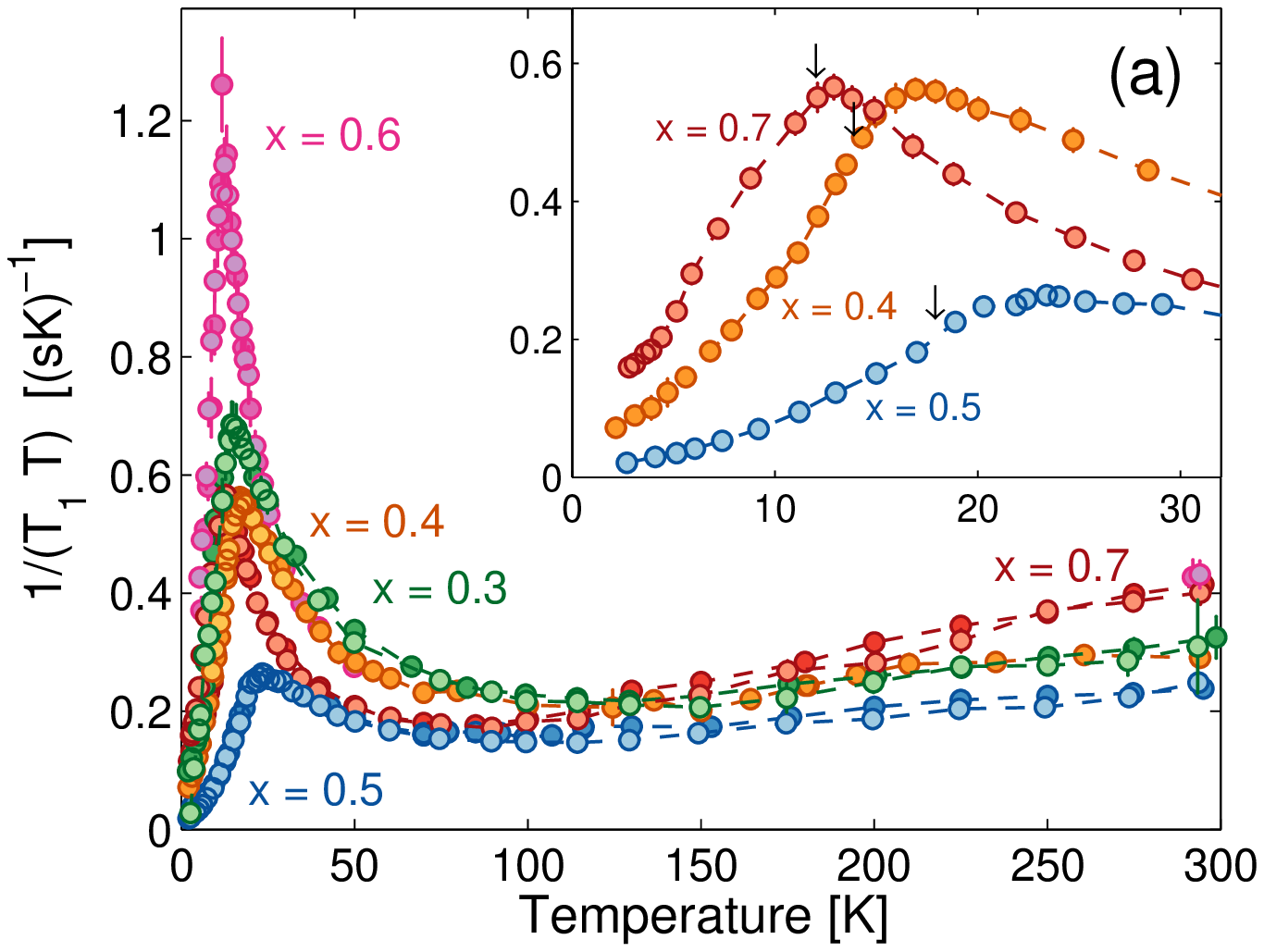} 
\hspace*{4mm}\includegraphics[width=0.9\columnwidth]{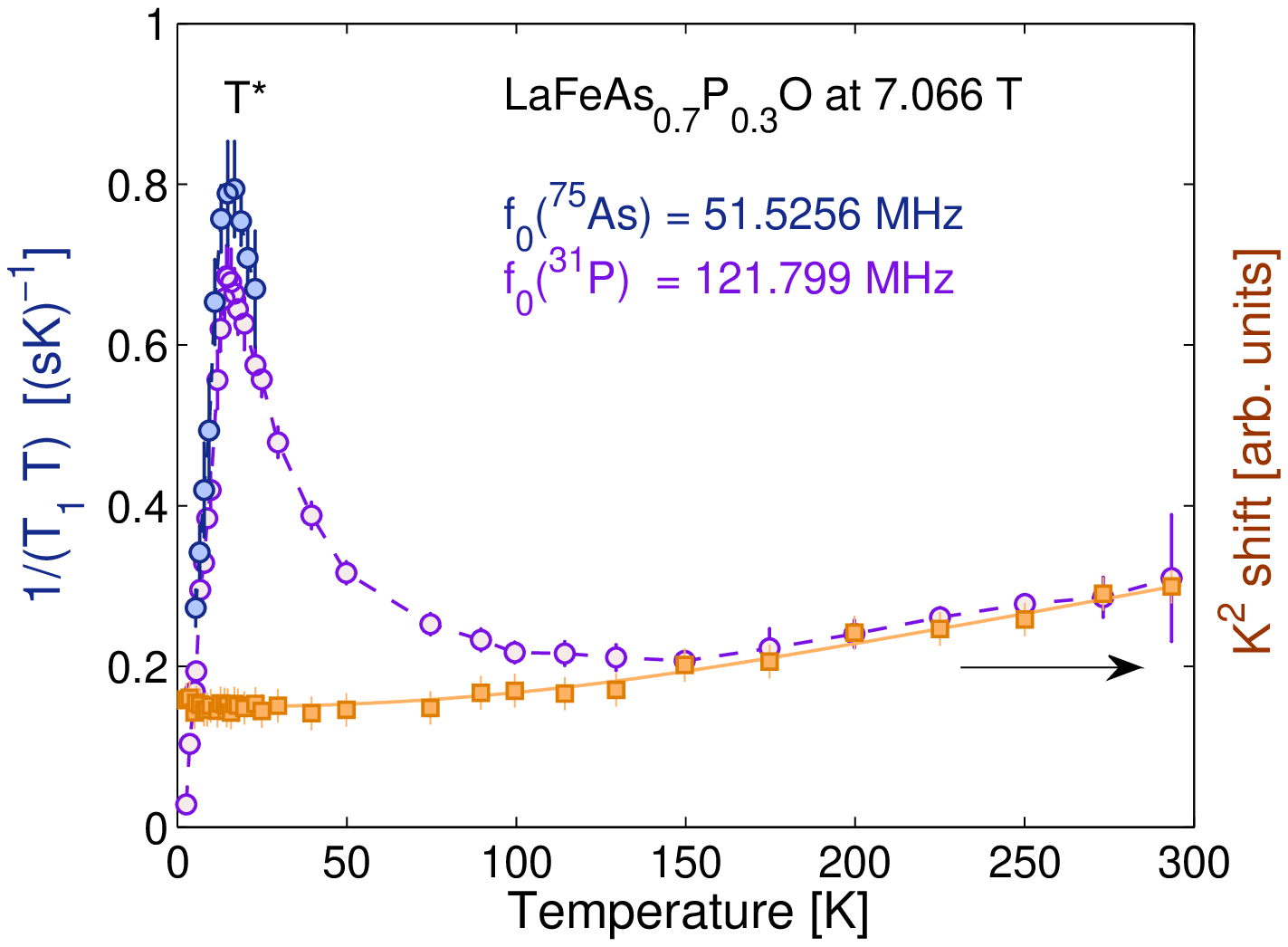} 
\caption{\textbf{Peak in $1/\boldsymbol{T_1T}$ hints at significant spin fluctuations, as 
confirmed by comparison with Knight-shift data.}
(a) $1/(T_1T)$ vs.\ temperature for all the samples. While the datasets at 3.5 
and 7\,T for the same sample coincide, the height and position of maxima depend 
strongly on $x$. 
Inset: Below $T_{c}$ (arrows) the lowest $1/(T_1T)$ value is achieved for the 
sample with the highest $T_{c}$ ($x = 0.5$). 
(b) Left scale (circles): $1/(T_1T)$ vs.\ temperature in LaFeAs$_{0.7}$P$_{0.3}$O 
measured via ${}^{31}$P and ${}^{75}$As NMR at 7.066\,T. In spite of the very different 
Larmor frequencies, the comparison shows closely matching features and similar $1/(T_1T)$ 
magnitudes. Given the absence of quadrupole effects for the $I = 1/2$ ${}^{31}$P 
nucleus, this similarity hints at a magnetic origin of nuclear relaxation, i.e., related 
to AF electron-spin fluctuations. 
Right scale: A comparison of the temperature dependences of $1/(T_1T)$ (circles) 
and ${}^{31}K^2$ (squares). The significant departure of the two curves below 
ca.\ 120\,K indicates the development of strong AF fluctuations.
}
\label{fig:invT1T_shift}
\end{figure*}

Here $M_z^0$ represents the saturation value of magnetization at thermal 
equilibrium, $f$ is the inversion factor (exactly 2 for a complete inversion), 
and $\beta$ is a stretching exponent. The latter is required, since for 
samples with intrinsic disorder multiple relaxation times are expected. 
Indeed, as shown in the inset of Fig.~\ref{fig:T1_curves_2T}, the recovery 
occurs over many decades, reflecting a wide distribution of relaxation rates. 

The evolution of $\beta$ with temperature indicates a smooth decrease from 1, 
the canonical value for simple disorder-free metals, to almost 0.5 close to $T = 0$\, K. 
Such a strong reduction of $\beta$ is typical of samples with disorder, where 
the inequivalence of NMR sites increases as the temperature is lowered\citens{Shiroka2011a}. 
As shown in Fig.~\ref{fig:T1_curves_2T}, samples having the same degree 
of disorder exhibit a very similar $\beta(T)$ dependence. Yet, the vertical 
offset, most likely indicates again a different degree of electronic 
correlations above and below $x = 0.5$.

Figure~\ref{fig:invT1T_shift}a summarizes the extensive
$1/(T_{1}T)$ dataset, collected at both fields and for all the samples. 
Unlike the Knight-shift and linewidth data, the $1/(T_{1}T)$ vs.\ $T$ curves 
are practically independent of the applied field for all the investigated $x$ values. 
We recall that $1/(T_{1}T) = \sum_{q}F(q)\chi''(q,f_\mathrm{r})/f_\mathrm{r}$ 
probes the fluctuating hyperfine fields at a nuclear site and, as such, 
it represents a measure of the dynamic correlations. Here, $F$ is the tensor of the 
hyperfine form-factor, while $\chi''$ represents the imaginary part of 
the dynamical electronic susceptibility. 
The main feature of the reported $(T_{1}T)^{-1}(T)$ data is the presence 
of low-temperature peaks of varying magnitude.  
The substantial increase of $(T_{1}T)^{-1}$ upon lowering the
temperature indicates an increase in the dynamical susceptibility, 
typical of a magnetic instability and/or spin fluctuations\citens{Oka2012}.
The successive steep decrease upon further cooling suggests instead a 
progressive slowing down of spin fluctuations, associated to a short-range 
diffusive dynamics in the MHz range, involving 
wall motions of nematic domains\citens{Hammerath2013,Bossoni2016}.
Since such a slow dynamics cannot be captured by faster techniques 
such as $\mu$SR, a much less pronounced peak is observed in the 
ZF-$\mu$SR relaxation rates (see Fig.~\ref{fig:ZF-MuSR}b).
This is further confirmed by the prompt decoupling of muon spins 
in longitudinal-field $\mu$SR measurements (not shown). 

By comparing the $T_{c}$ values vs.\ $x$ (as determined via susceptibility 
measurements --- see Supplementary Fig.~\ref{fig:magnetization}) we note 
that the sample with the highest $T_{c}$ does not display 
the most intense spin fluctuations, but rather the opposite is true. 
The complete set of $1/(T_{1}T)$ data shows that, as in 
case of Knight shifts, samples with $x$ values above and below 
$x = 0.5$ do not exhibit the same relaxation curves. This persistent 
lack of symmetry indicates a significant change in the electronic properties 
of the \LOFP\ series across the $x = 0.5$ demarcation line.

Further insight into the electronic correlations and spin fluctuations 
across the \LOFP\ series is obtained from two instructive comparisons, 
both presented in Fig.~\ref{fig:invT1T_shift}b. First, we compare 
the $K^2(T)$ behavior with the temperature dependence of $1/(T_1T)$.
Since in simple metals, both the Knight shift and the relaxation rate depend 
essentially on the electronic density of states at the Fermi level, 
$N(E_\mathrm{F})$, the two curves should adopt a similar functional form, 
as expected from the Korringa relation $K^2 = S \cdot 1/(T_1T)$,  
with $S$ a constant\citens{Korringa1950}. 
The Knight shift probes only the uniform susceptibility, whereas 
$1/(T_1T)$ depends also on the electron-spin dynamics. A clear departure of 
the two, as observed in our case below 90\,K, indicates the development 
of significant antiferromagnetic spin fluctuations. 
The peak in $1/(T_1T)$ correlates with the onset of an NMR line broadening 
(see Fig.~\ref{fig:shifts}b), which at first might suggests the onset of an 
AF order. However, the tiny increase in FWHM and the practically 
constant $\mu$SR relaxation with temperature (see Fig.~\ref{fig:ZF-MuSR}), 
both rule out the occurrence of a proper magnetic order, indicating instead 
a spin-fluctuation dominated scenario, with the opening of a spin-gap below $T^{*}$.

The spin-fluctuation driven relaxation is confirmed also by a second comparison, 
that of the ${}^{31}$P and ${}^{75}$As NMR relaxation rates. 
Both of them are plotted in Fig.~\ref{fig:invT1T_shift}b as $1/(T_1T)$ 
vs.\ $T$ for the $x = 0.3$ case. Although the resonance frequencies differ by 
more than a factor of 2, the two datasets almost coincide. 
This is true not only for the position of the  $1/(T_1T)$ peaks, but 
suprisingly also with regard to the almost equal magnitudes. 
Since the two nuclei have spins $I = 1/2$ and $3/2$, they can relax by 
means of magnetic-only and magnetic and quadrupole relaxation channels, 
respectively. 
\begin{figure}[ht!]
\includegraphics[width=0.85\columnwidth]{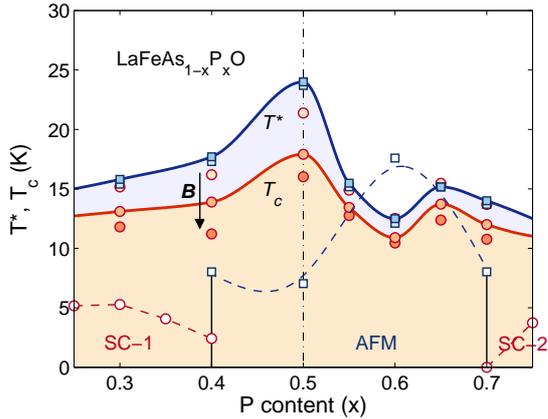} 
\caption{\label{fig:phase_diag}Phase diagram of \LOFP\ showing the critical temperature 
$T_c$ and that of the spin-fluctuation maxima $T^*$ at different applied fields 
(vertical arrow: 0, 3.5, and 7\,T), measured via magnetometry and NMR, respectively. 
While the onset of superconductivity is suppressed by the applied field, 
the $T^*$ values remain unaffected. 
Empty symbols and dashed lines refer to the phase diagram of the material 
grown at ambient pressure\citens{Mukuda2014}, exhibiting two SC phases 
separated by an antiferromagentic phase. For clarity, the latter temperature values 
were reduced by half.
}
\end{figure}
The practically overlapping $1/(T_1T)$ peaks indicate that quadrupole effects 
play no (or only a minor) role in the relaxation of ${}^{75}$As nuclei. Therefore, 
the only remaining relaxation channel, available in both cases, is that dominated 
by magnetic interactions, which in our case can be identified with spin fluctuations.

\section{Discussion}\label{sec:discussion}
To summarize the results of the different measurements on the \LOFP\ 
series reported above, we provide an overview in the form of the phase 
diagram shown in Fig.~\ref{fig:phase_diag}. We notice that: (a) $T_c$ 
reaches a maximum for $x = 0.5$, (b) the diagram is not symmetric with 
respect to this value, and (c) the phase diagram is very different from 
that of samples grown at ambient pressure.

The reason for the maximum 
$T_{c}$ being reached for $x = 0.5$ is most likely related to the 
pnictogen-height value $h_\mathrm{Pn}$ over the iron plane. 
Detailed structural analyses of a similarly synthesized 1111 family 
with isolectronic pnictogen substitution have shown that the $x = 0.5$ 
composition corresponds to the highest $T_c$ and to 
$h_\mathrm{Pn} = 1.32$\,\AA\citens{Zhigadlo2011}. 
The latter is very close to the optimal $h^\mathrm{opt}_\mathrm{Pn} = 1.38$\,\AA\ 
value, known to produce the highest $T_c$s in many classes of iron-based 
superconductors\citens{Mizuguchi2010}.
On the theoretical side, models of superconductivity based on a spin-fluctuation 
mediated pairing correlate $h^\mathrm{opt}_\mathrm{Pn}$ with the 
electron-hole interband scattering rate (see, e.g., Ref.~\onlinecite{Thomale2011}), 
with the optimum value achieved exactly in the symmetric $x = 0.5$ 
case. In our case this would imply that, in spite of a spin-gap 
opening below $T^{*} (> T_{c})$, it would still allow for the 
formation of a superconducting state below $T_{c}$.

The phase diagram asymmetry, instead, may reflect a symmetry change in 
the superconducting order parameter, from \textit{nodal to nodeless}, 
when $x$ increases from 0 in LaFeAsO to 1 in LaFePO. 
Indeed, it has been pointed out that the transition between the two different types 
of SC order parameter occurs at $h_\mathrm{Pn} = 1.33$\AA,\citens{Kuroki2009,Hashimoto2012} 
practically coincident with the $h^\mathrm{opt}_\mathrm{Pn}$ value reported above, 
although the change in SC character is opposite in our case, probably due to the 
high-pressure synthesis conditions.
Since $h^\mathrm{opt}_\mathrm{Pn}$ corresponds to $x = 0.5$ in our case, 
this implies that compounds such as LaFeAs$_{0.6}$P$_{0.4}$O and 
LaFeAs$_{0.4}$P$_{0.6}$O, deviating by $\pm 0.1$ from $x = 0.5$, should 
behave differently. 
Indeed, the data reported above show clear variations across $x = 0.5$ 
in the temperature dependences of resistivities, $K$-shift values, and $1/(T_1T)$ 
rates, as well as in the TF-$\mu$SR parameters.
Our results, therefore, provide strong support in favor of 
$h_\mathrm{Pn}$ acting as a \emph{switch}
between the nodal and nodeless pairings\citens{Kuroki2009},  
with $h_\mathrm{Pn}$ being determined by the As-to-P substitution ratio. 
Ultimately, it is the change in the lattice structure which modifies the 
nesting among disconnected parts of the Fermi surface (FS).  
This makes the Fermi-surface topology one of the key parameters to 
determine the occurrence of superconductivity, whereas the exchange 
interaction between localized  Fe$^{2+}$ moments in the 3$d$ orbitals 
is the other one\citens{Arita2009,Usui2015}.

The above mentioned orbital effects are crucial to understand why a maximum 
$T_c$ is achieved at intermediate $x$ values ($x = 0.5$, in our case). Upon 
increasing the As/P ratio, the hybridization between the $d_{XZ}$ and $d_{YZ}$ 
orbitals (a 45-degree rotated version of the standard $d_{xz}$ and $d_{yz}$ orbitals)  
is enhanced\citens{Usui2015}.
On the one hand, hybridization optimizes the \textit{orbital matching} between the 
electron- and hole Fermi surfaces and enhances the spin fluctuations within the orbitals, 
in turn acting as mediators of the superconductivity. 
An increased hybridization also decreases the intersection of the two relevant 
ellipse-shaped Fermi surfaces, generating a favorable nesting for superconductivity. 
On the other hand, the hybridization splits the two bands, with the more 
dispersive inner band achieving a lower density of states, thus implying 
lower $T_c$ values. The final outcome of these opposing trends upon isoelectronic 
doping is a \textit{compromise} between orbital matching and a reduction 
in the density of states, which results in an optimal $T_c$ 
at intermediate As/P ratios, as observed experimentally.

Finally, we emphasize that a phase diagram, where superconductivity is 
found for all the $x$ values between 0.3 and 0.7, is very different from 
the multi-dome diagram found for samples synthesized at ambient 
pressure\citens{Mukuda2014,Miyasaka2017}. 
Since quenching is known to stabilize otherwise metastable states  
obtained under high-pressure high-temperature conditions, this can 
explain the essential differences observed in the two cases.

In conclusion, by using different micro- and macroscopic techniques, 
we investigated the electronic properties of the \LOFP\ family of 1111 
iron-based superconductors. 
Our results, show that samples from the same family when synthesized under 
high-pressure, differ in fundamental ways from those synthesized under 
ambient-pressure conditions. 
Our key finding, supported by both ZF-$\mu$SR and NMR results, is the 
\emph{lack of antiferromagnetic order} in all the compounds covered in our 
investigation. Instead, we find clear evidence of \emph{significant spin fluctuations} 
across the $0.3 \leq x \leq 0.7$ range of the series.
In addition, unlike in the previously reported results, we find an onset of 
superconductivity for all our samples, with $T_c$ values depending on $x$, 
lying at or slightly below the temperatures where relaxation rates due to spin 
fluctuations reach their maxima. This proximity suggests a close competition between 
the incipient magnetic order and superconductivity, with the latter most likely being 
mediated by spin fluctuations.
Finally, the asymmetric character of the \LOFP\ phase diagram, as well as 
the distinctly different NMR datasets for samples with nominally symmetric 
compositions with respect to $x = 0.5$, indicate the different nature of the 
superconducting order parameter across the $x = 0.5$ boundary, 
evolving from nodal to nodeless as $x$ increases.
The peculiar behavior of La-1111 grown under high pressure conditions, 
implies that even \textit{nominally identical} As concentrations can produce 
very different local environments and, therefore, give rise to a different evolution 
of $T_{c}$ as $h_\mathrm{Pn}$ is modified via chemical substitution\citens{Zhigadlo2011}.
In view of this, other high-pressure grown iron-based superconductors are expected 
to be in for new surprises.

\vspace{5mm}

\section{Methods}
\subsection{Sample preparation and characterization.}
\footnotesize{%
A series of polycrystalline \LOFP\ samples was prepared by using the 
cubic-anvil high-pressure and high-temperature 
technique\citens{Zhigadlo2012,Zhigadlo2013,Zhigadlo2016}.
Due to the toxicity of arsenic, all procedures related to the sample preparation 
were performed in a glove box. Pellets containing the high-purity ($>99.95$\%) 
precursors (La$_2$As, LaP$_2$, Fe$_2$O$_3$, As, and Fe) were enclosed in 
a boron nitride container and placed into a graphite heater. A pressure 
of 3\,GPa was applied at room temperature. Then, by keeping the pressure 
constant, the temperature was ramped up to 1320$^{\circ}$C in 2\,h, maintained 
there for 12\,h, and finally abruptly quenched to room temperature. 
Once the pressure was released, the sample was removed. 
The structural characterization was performed by means of standard powder 
x-ray diffraction (XRD) measurements carried out at room temperature, 
which confirmed the single-phase nature of the samples, as well as the 
absence of impurities (below the 1\% level). 
Temperature-dependent DC magnetization measurements were performed by 
means of a superconducting quantum interference device (SQUID) magnetometer 
(Quantum Design), while the electrical resistivity of pressed powder 
specimens was measured in a four-point probe configuration. Finally, 
energy-dispersive x-ray (EDX) spectroscopy was used to quantitatively 
analyze the chemical composition of the synthesized samples.
}

\subsection{NMR and $\mu$SR measurements.}
\footnotesize{%
For the microscopic investigation of \LOFP, with $0.3 \leq x \leq 0.7$, 
in both the normal and the superconducting phase, we employed first  
${}^{31}$P NMR. With an isotopic abundance of 100\% and a high gyromagnetic 
ratio ($\gamma/2\pi = 17.254$\,MHz/T), this $I = 1/2$ nucleus provides 
a favorable local probe. In selected cases we also performed ${}^{75}$As 
NMR measurements. A good signal-to-noise (S/N) ratio was achieved by 
using samples in the form of loose powders, which reduces the electrical 
contacts between grains. The NMR spectra in the 2--300\,K range were 
obtained by fast Fourier transformation (FFT) of the spin-echo signals 
generated by $\pi/2 - \pi$ rf pulses with 50\,$\mu$s of typical delay 
between the pulses. 
Given the short rf pulse length ($t_{\pi/2} \sim 3$ $\mu$s), frequency 
sweeps were not necessary for acquiring the $^{31}$P NMR lines. 
Since samples with intermediate $x$ contain two independent 
NMR-active nuclei, in selected cases we also performed ${}^{75}$As-NMR measurements. 
Given the nuclear spin $I = 3/2$ and related quadrupole effects for ${}^{75}$As, 
this allows for an instructive comparison with the purely-magnetic spin-$1/2$ 
${}^{31}$P data (see above).
In addition, we investigated the effects of the applied magnetic field, 
by acquiring NMR data at $\mu_0H = 7.066$\,T and 3.505\,T.
Nuclear spin-lattice relaxation times $T_1$ were measured following a 
standard inversion-recovery procedure with spin-echo detection at variable delays.
The magnetic field was calibrated using ${}^{27}$Al NMR on pure aluminum,
whose gyromagnetic ratio and Knight shift are known to high precision.

The $\mu$SR measurements were performed at the general-purpose spectrometer (GPS) 
of Paul Scherrer Institut, PSI, Villigen (Switzerland). 
Various powder samples from the \LOFP\ series were mounted on copper forks by using 
aluminated mylar and kapton foils. This setup up, combined with active vetoing, resulted 
in very low spurious background signals. 
Due to active compensation coils, true zero-field conditions were achieved during the 
ZF-$\mu$SR experiments. 
The ZF and TF-$\mu$SR measurements were carried out between 1.5 and 30\,K, 
the lowest temperatures being reached by using a pumped He-4 cryostat. 

The error bars in case of $\mu$SR measurements were obtained from the raw 
data counting statistics, while for the NMR they were derived from the NMR-signal 
noise levels. The reported error bars were calculated by using the standard 
methods of error propagation.
}

\subsection{\label{app:Fit_formulas}Fitting formulae for the superconducting gap.}
\footnotesize{%
TF-$\mu$SR measurements give access to $\lambda^{-2}(T)$, which is 
proportional to the effective superfluid density, $\rho_s \propto \lambda^{-2}$.
Hence, a study of the temperature dependence of  $\lambda^{-2}(T)$ can reveal 
the symmetry of the superconducting gap (i.e., of the electronic 
density of states in the proximity of the Fermi energy below $T_c$).
As shown in Fig.~\ref{fig:TF-MuSR}c (solid dark line), the experimental 
$\lambda^{-2}(T)$ data for $x > 0.5$ are consistent with a \textit{nodeless} superconducting 
gap with $s$-wave symmetry, which in the clean limit regime ($l > \xi$) gives\citens{Tinkham1996}: 
 \begin{equation}\label{eq:fit_lam_clean}
\left[ \frac{\lambda(0)}{\lambda(T)}\right]^{2} = 1 + 2 \int_{\Delta(T)}^{\infty}\! 
\left(\frac{\partial f}{\partial E}\right) \frac{E}{[E^2-\Delta(T)^2]^{1/2}}\, \mathrm{d}E.
\end{equation}
Here  $\lambda^{-2}(0)$ is the zero-temperature value of the magnetic penetration depth 
and $f=[1+\exp(E/k_\mathrm{B}T)]^{-1}$  represents the Fermi distribution. 
The temperature dependence of the superconducting gap can be approximated analytically as\citens{Carrington2003}: 
 \begin{equation}\label{eq:Delta_T_approx}
\Delta(T) = \Delta_0 \tanh \left\{ 1.82 \left[ 1.018 \left( \frac{T_c}{T}-1 \right) \right]{}^{0.51} \right\},
\end{equation}
with $\Delta_0$ the gap value at zero temperature.

In the $x < 0.5$ case, however, the nodeless $s$-wave model in 
Eq.~(\ref{eq:fit_lam_clean}) cannot fit the data (see Fig.~\ref{fig:TF-MuSR}b, 
solid gray line). Only a $d$-wave based model, which contains \textit{nodes}, 
can account for the experimental  $\lambda^{-2}(T)$ data. In this case the 
superconducting gap $\Delta = \Delta(T,\phi)$ acquires an additional 
 $\left| \cos(2\phi)\right|$ angular factor and the temperature dependence 
 of $\lambda^{-2}(T)$ becomes:
 \begin{equation}\label{eq:fit_d_wave}
\left[ \frac{\lambda(0)}{\lambda}\right]^{2}=1 + \frac{8}{\pi} 
\int_{0}^{\frac{\pi}{4}}\!\!\int_{\Delta}^{\infty}\!\!
\left(\frac{\partial  f}{\partial  E}\right) \frac{E}{[E^2-\Delta^2]^{1/2}}\, \mathrm{d}E \mathrm{d}\phi.
\end{equation}
The fits with an $s$-wave model for $x > 0.5$ and a $d$-wave model for $x < 0.5$, 
give $\lambda(0) = 391(10)$\,nm and $\Delta_{s}(0) = 2.51(2)$\,meV for $x = 0.65$ 
and $\lambda(0) = 476(10)$\,nm and $\Delta_{d}(0) = 3.10(3)$\,meV for $x = 0.4$.
Considering the similar $T_c$ values, the $2\Delta(0)/k_{\mathrm{B}}T_{c}$ ratios 
are 3.6 and 5.3, respectively, to be compared with 3.52 of the standard BCS theory.
}

\subsection{Data availability.} 
\footnotesize{%
The data that support the findings of this study are available 
from the corresponding author upon reasonable request.
}

\section{Acknowledgments} 
\footnotesize{%
The authors thank A.\ Amato (Paul Scherrer Institut) for
the assistance during the experiments and P.\ Macchi for useful discussions.
This work was financially supported in part by the Schweizerische 
Nationalfonds zur F\"{o}rderung der Wissenschaftlichen Forschung (SNF). 
}

\section{Author contributions}
\footnotesize{%
Project planning: T.S. Sample synthesis and characterization: N.D.Z. $\mu$SR 
experiments were carried out by R.K.; NMR measurements and data analysis 
by T.S.\ and N.B. The manuscript was drafted by T.S.\ and H.R.O.\ and was 
completed with input from all the authors.
}

\footnotesize{%
\vspace{2mm}
\noindent\textbf{Competing financial interests:} The authors declare 
no competing financial interests.
}

\vspace{-5mm}
\renewcommand*{\bibfont}{\footnotesize}

%

\cleardoublepage{}

\renewcommand{\figurename}{Supplementary Figure}
\renewcommand{\tablename}{Supplementary Table}
\makeatletter\renewcommand{\fnum@figure}[1]{\textbf{\figurename~\thefigure~|\ }}\makeatother
\makeatletter\renewcommand{\fnum@table}[1]{\tablename~\thetable.}\makeatother

\section{Supplementary information}
\setcounter{figure}{0}    
\normalsize{%
The detailed x-ray diffraction patterns are shown in Fig.~\ref{fig:diffraction}. Here the 
main peaks refer to the diffraction from the LaO and FeAs(P) planes, while the highlighted 
area indicates the minor peaks corresponding to reflections from the (110) 
and (003) planes, which relate to the local arrangement of atoms. The latter 
differs significantly from that of samples grown at ambient pressure, hence justifying 
the rather different properties of samples grown under high-pressure conditions.

\begin{figure}[!htb]
\centering
\includegraphics[width=0.79\columnwidth]{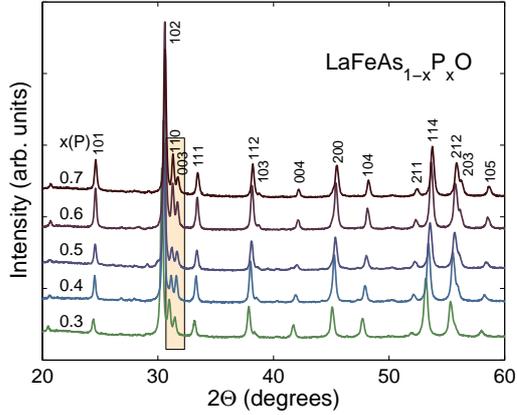} 
\caption{\label{fig:diffraction}\textbf{Powder diffraction patterns.}
The room-temperature x-ray powder diffraction patterns of \LOFP\ exhibit 
a regular evolution with $x$ and show no traces of spurious phases.
The multiple peaks close to 30 degrees (highlighted area) evolve differently 
from those in samples grown under ambient pressure, indicating 
a different local environment within the FeAs planes.
}
\end{figure}

Data on magnetization are reported in Fig.~\ref{fig:magnetization}.
Due to a rather high estimated $H_{c2}(0)$ value of ca.\ 70\,T, the 
applied magnetic fields chosen for the NMR measurements do not induce 
a significant lowering of $T_c$, with both magnetometry and in-situ RF 
detuning showing a shift in $T_c$ of ca.\ $-2.5$\,K at 7\,T (see inset).

\begin{figure}[!htb]
\centering
\includegraphics[width=0.85\columnwidth]{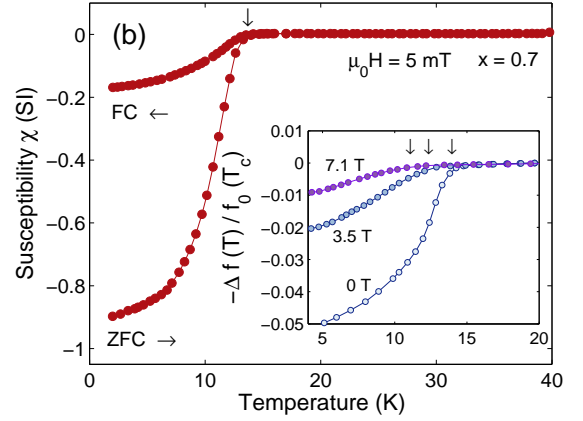}
\caption{\label{fig:magnetization}\textbf{Magnetic susceptibility data.} 
Zero field-cooled (ZFC) and field-cooled (FC) dc susceptibility vs.\ temperature 
measured at 5\,mT in the $x = 0.7$ case. Inset: tank-circuit detuning vs.\ $T$ 
at different applied fields was used to determine $T_c(H)$. Arrows denote 
the $T_c$ positions.
}
\end{figure}

From the analogous magnetization data for the rest of the investigated 
samples, shown in Fig.~\ref{fig:magn_all}, we determine the relevant 
critical $T_{c}$ values, as reported in Fig.~\ref{fig:phase_diag}.
\begin{figure}[!htb]
\centering
\includegraphics[width=0.85\columnwidth]{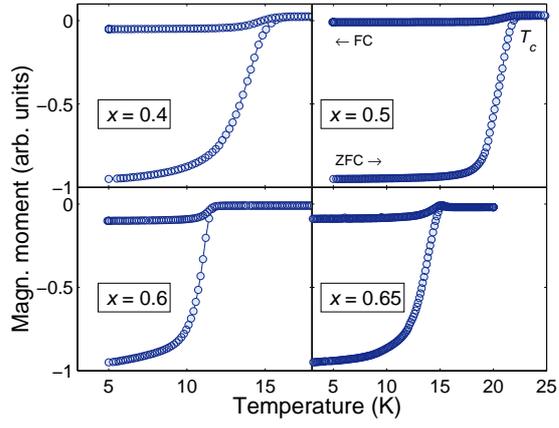}
\caption{\label{fig:magn_all}\textbf{Magnetization data for all the samples.} 
Zero field-cooled (ZFC) and field-cooled (FC) dc magnetization vs.\ temperature 
measured in fields of 3 to 10\,mT for the $x = 0.4$, 0.5, 0.6, and 0.65 case. 
The FC and ZFC curves, as well as the relevant $T_{c}$ are indicated in 
the $x = 0.5$ panel.
}
\end{figure}

\begin{figure}[!htb]
\centering
\includegraphics[width=0.60\columnwidth]{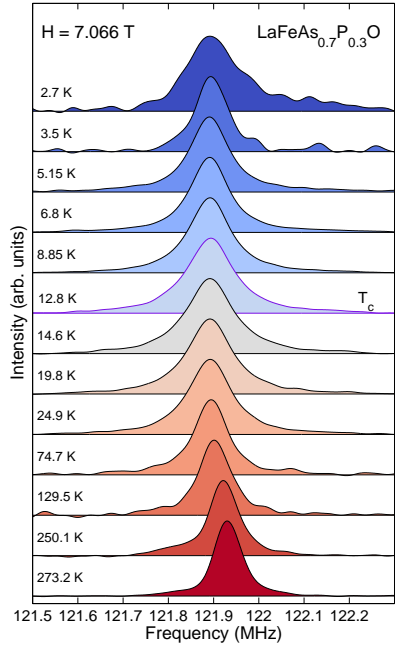}
\caption{Representative ${}^{31}$P NMR line shapes in LaFeAs$_{1-x}$P$_x$O 
(for $x = 0.3$) at $\mu_0H = 7.066$\,T and temperatures in the 2.7 to 270\,K 
range. The increased line width below $T_c = 12$\,K reflects the onset of 
the superconducting phase.}
\label{fig:lines}
\end{figure}
} 

\end{document}